\documentclass[preprint,showpacs,preprintnumbers,amsmath,amssymb,prb]{revtex4}
\usepackage{graphicx}
\usepackage{dcolumn}
\usepackage{bm}

\begin{document}

\title{ 
Large Attractive Depletion Interactions \\ 
in Soft Repulsive--Sphere Binary 
Mixtures
} 

\author{Giorgio Cinacchi}
\email{g.cinacchi@sns.it}
\homepage{http://www.dcci.unipi.it/~ivo/cinacchi.htm}
\affiliation{Dipartimento di Chimica e Chim. Ind. , Universit\`{a} di Pisa,
Via Risorgimento 35, I--56126 Pisa, ITALY}
\author{Yuri Mart\'{\i}nez--Rat\'{o}n}
\affiliation{Grupo Interdisciplinar de Sistemas Complejos (GISC), Departamento de Matem\'{a}ticas,
Escuela Polit\'{e}cnica Superior, Universidad Carlos III de Madrid,
Avenida de la Universidad 30, E--28911 Legan\'{e}s, Madrid, SPAIN}
\author{Luis Mederos}
\affiliation{Instituto de Ciencia de Materiales de Madrid, Consejo Superior de Investigaciones Cient\'{\i}ficas, E-28049 Cantoblanco, Madrid, SPAIN}
\author{Guillermo Navascu\'{e}s}
\affiliation{D{\it{e}}partamento de F\'{\i}sica T\'{e}orica de la Materia Condensada,
 Universidad Aut\'{o}noma de Madrid,
E--28049 Madrid, SPAIN}
\author{Alessandro Tani}
\affiliation{
Dipartimento di Chimica e Chim. Ind., Universit\`{a} di Pisa,
Via Risorgimento 35, I--56126 Pisa, ITALY}
\author{Enrique Velasco}
\affiliation{Departamento de F\'{\i}sica T\'{e}orica de la Materia Condensada
and Instituto de Ciencia de Materiales Nicol\'as Cabrera,
 Universidad Aut\'{o}noma de Madrid,
E--28049 Madrid, SPAIN}

\begin{abstract}
\noindent We consider binary mixtures of soft repulsive spherical particles
and calculate the depletion interaction between two big spheres mediated by 
the fluid of small spheres, using different theoretical and simulation methods. 
The validity of the theoretical approach, a virial expansion in terms of the
density of the small spheres, is checked against simulation results. 
Attention is given to the approach toward the hard--sphere limit, and to the 
effect of density and temperature on the strength of the depletion potential. 
Our results indicate, surprisingly, that even a modest degree of softness in the 
pair potential governing the direct interactions between  the particles may lead 
to a significantly more attractive total effective potential for the big 
spheres than in the hard--sphere case.
This might lead to significant 
differences in phase behavior, structure and dynamics of a binary mixture of 
soft repulsive spheres.
In particular, a perturbative scheme is
applied to predict the phase diagram of an effective system of big spheres 
interacting via depletion forces for a size ratio of small and big
spheres of $0.2$; this diagram includes the usual
fluid--solid transition but, in the soft--sphere case, the metastable 
fluid-fluid transition, 
which is probably absent in hard--sphere mixtures, 
is close to 
being stable with respect to direct fluid--solid coexistence. From these 
results the interesting possibility arises that, for sufficiently soft 
repulsive particles, this phase transition could become stable.
Possible implications for the phase behavior of real 
colloidal dispersions are discussed.
\end{abstract}

\pacs{82.70.Dd} 

\maketitle
\clearpage

\section{Introduction }

Colloidal systems are complex fluids composed of mesoscopic particles dispersed in a solvent
of microscopic particles. The large difference in size between the 
dispersed (colloidal) and the solvent entities may allow the latter to be considered
as a continuum \cite{hunter}.

One way to prevent colloidal particles from aggregating is to sterically stabilize them,
i.e. to coat them with a suitable polymer layer. This leads to an effective repulsive 
interaction, frequently considered so steep that 
colloidal particles have been extensively modeled as  hard  bodies. 
Thus, a dispersion made of spherical colloidal particles may in a first approximation be assimilated
to a system of hard spheres (HS), the phase behavior of which is well known. In this respect, 
systems of mono--sized polymethylmethacrylate particles provide a good example
\cite{pusey86}. 

The case of binary colloidal dispersions, where two colloidal species of 
different size are immersed in a molecular solvent, is less clear. Here the HS model may
also be used to characterize colloidal interactions. Sophisticated integral--equation theories
for binary HS fluid mixtures indicate that, contrary to the prediction of the 
Percus--Yevick closure \cite{lebowitz}, demixing between two phases of different 
compositions may occur, provided the diameters of the two components are sufficiently 
different \cite{biben}. Binary mixtures of silica particles were shown to undergo
phase--separation \cite{lekker93}, but its nature was not ascertained. The possibility that
the phases involved were both fluid could not be excluded, but recent work has
given evidence that demixing involves fluid and solid phases, and that fluid--fluid
segregation in HS mixtures should be excluded \cite{HSs,dvre,dvre1,Evans}.

The mechanism supposed to be at the basis of these demixing phenomena is known as the
\emph{depletion effect}, according to which there exists an effective \emph{attractive} 
interaction between the big particles due to their clustering leading to a larger free 
volume available for the small particles. It was first put forward by
Asakura and Oosawa \cite{AO}, and later reconsidered by Vrij \cite{Vrij},
to explain the phase behavior of colloid--polymer 
mixtures. Depletion effects are invoked to explain a large variety of phenomena
in mixtures and have been the subject of intense theoretical effort in recent years.
The validity of the effective one--component approximation (big colloidal particles 
interacting via a depletion potential) in describing the phase behavior of the actual
binary HS mixture has been checked by comparison of the phase behavior
obtained from the effective potential of mean force using computer simulation, with that 
obtained from direct computer simulations of the mixture \cite{dvre,dvre1}.
These studies revealed that fluid--fluid phase separation is actually unstable with respect 
to a fluid--solid phase transition; other computer simulation studies, based again on the 
effective one--component approach, showed no indication of any (even metastable) 
fluid--fluid demixing \cite{almarza}, suggesting that slightly different approximations 
for the depletion potential employed may lead to quite different phase behavior.

The very shape of the depletion potential appears to be very sensitive to small variations
in the direct interactions among particles. For instance, slight degrees of 
non--additivity in binary HS systems result in depletion potentials which are 
very different from those calculated for the corresponding additive binary mixtures 
\cite{nonaddit}. Also, residual interactions of short range and moderate strength, such as 
very short--ranged {\it attractive} Yukawa tails, which {\it prima facie} 
would appear irrelevant 
compared to the dominant hard--body terms, have been shown to significantly affect phase
behavior \cite{amokrane1}, tending to make the effective interactions {\it less} attractive.
Moreover, the 
importance of carefully taking into account the colloidal--colloidal direct
interactions has been stressed \cite{amokrane2}: the total effective
potential is the sum of the direct and indirect (depletion) contributions,
and clearly the final shape depends on very fine details of both.

The above--mentioned sensitivity of (and on) the shape of the depletion potential undermines
the rather common view that mixtures of sterically--stabilized spherical colloidal particles
resemble HS mixtures, and serves as a stimulus for further investigations. 
In particular, a largely ignored, and probably more realistic, variation on 
the hard--sphere theme is to consider particles to have soft, rather than hard, repulsive 
cores. Although quite convincing experimental evidence exists about the fact that colloidal 
particles coated with poly--12--hydroxystearic acid, one of the most frequently used 
coating agents, should really behave as hard spheres in many solvents \cite{australia}, 
it appears of relevance to investigate depletion interactions in {\it repulsive} particles
of various degrees of softness. In fact, the softness of the interaction potential was shown
to have a pronounced influence on the crystal nucleation of model colloidal systems 
\cite{auer}, while interactions between colloidal particles may be tuned to a certain degree,
for example by varying the chemical composition and thickness of the surface layer,
to such an extent that soft--core colloidal particles can be realized \cite{duijneveldt} .

In this work, binary fluid mixtures of model soft repulsive spherical particles have
been considered. The depletion interaction between two big particles
induced by the small-particle component has been evaluated using computer simulation.
The depletion interaction can be obtained quite straightforwardly,
 using Molecular Dynamics (MD) simulation,
as the potential of mean force on a big sphere. Systems of 
various densities have been analyzed and the results compared with various approximations
based on a virial expansion. The most salient feature of the resulting depletion potential
is that the effective attraction between soft spheres is greatly enhanced with respect to
the HS case: as the softness is increased, the attraction becomes substantially
larger, and the effective diameter of the big spheres is reduced. Since these trends of
the interaction potential tend to promote fluid--fluid separation in liquids, the 
interesting possibility arises that fluid--fluid demixing could be stabilized in these fluids
of repulsive particles. In an effort to explore this possibility, we have 
applied a perturbative scheme to evaluate the free energy, and from there
the complete phase diagram, for a size ratio of 
small and big spheres of 
$0.2$. 
These results already indicate that fluid--fluid demixing, which a related
theoretical calculation predicts it to be absent in the case of HS mixtures, is close
to being stable with respect to direct fluid--solid phase separation,
suggesting that further variations of the interactions,
for example by further softening the interaction potentials, might induce
the stability of a liquid phase in mixtures of soft repulsive spheres. 

The article is organized as follows. In Sec.
\ref{I} interaction models, simulation techniques and theoretical approximations for the
depletion potential are described. Results are contained in Sec. \ref{Pres}, and the 
conclusions are presented in Sec. \ref{III}.

\section{Theoretical approach and simulation details}
\label{I}

The fluids considered in this work are binary mixtures whose
constituents are spherical particles of two species, $i$ and $j$, interacting 
by a pair potential $u_{ij}(r)$ defined as

\begin{equation}
u_{ij}\left(r\right)= \left\{
\begin{array}{ll}\displaystyle
4\epsilon_{ij}\left[\left(\frac{\sigma_{ij}}{r}\right)^{2n}-
\left(\frac{\sigma_{ij}}{r}\right)^n+\frac{1}{4}\right], &\quad r \leq 2^{1/n} \sigma_{ij}, \\
\\\displaystyle
0,& \quad r > 2^{1/n} \sigma_{ij},
\end{array}
\right.
\label{defpot}
\end{equation}
with $\epsilon_{ij}$ and $\sigma_{ij}$ the energy and range parameters, respectively,
and $r$ the distance between the two interacting particles. 
$n$, henceforth called `softness parameter', 
is an exponent such that the limit $n\to\infty$ gives the HS case.
The rate at which the HS limit is approached as $n\to\infty$
in $u_{ij}(r)$ is dictated by the temperature $T$.
As $n$ is decreased
from this limit the repulsive potential becomes softer.
For the case $n=6$ this potential corresponds to the repulsive part of the 
Weeks--Chandler--Andersen separation scheme for the Lennard--Jones potential \cite{wca}.
In this work, $\epsilon_{ij}$=$\epsilon$ $\forall i,j$,
while the size parameters for the two species have been given the following values:
$\sigma_{11}$=$\sigma$, $\sigma_{22}$=0.2$\sigma$, $\sigma_{12}$=$\sigma_{21}$=0.6$\sigma$
(here an additivity rule is applied on the $\sigma_{ij}$ parameters; in this way the
limit $n\to\infty$ exactly corresponds to additive hard spheres). Particles
of the species labeled as 1 will therefore be referred to as big, solute or
colloidal spheres, while particles labeled as 2 will be called small or solvent 
spheres.                

One can define a total interaction potential $U(r)$ between two isolated big 
spheres immersed in the fluid of small spheres. This potential will have
two contributions: the direct, or bare, contribution, given by 
$u_{11}(r)$, and an indirect contribution, $W(r)$, which is transmitted
through the small spheres and is actually the potential of mean force;
this contribution is identified as the depletion potential, and the
total potential will be calculated as $U(r)=u_{11}(r)+W(r)$.
One of the aims of this work is to study how the depletion interaction, and
consequently the total effective interaction between the two big particles
varies not only with the density of the solvent, $\rho_2$,
but also with the degree of softness, $n$, and the temperature $T$.

In the case of generic particle systems,
it is convenient to measure physical quantities in reduced units.
Thus, we will define reduced densities 
as $\rho_i^*=\rho_i\sigma_{ii}^3$, $i=1,2$,
while a reduced temperature will be given as
$T^*=k_{\rm B}T/\epsilon=(\beta\epsilon)^{-1}$, with $\beta=1/k_{\rm B}T$ and
$k_{\rm B}$ Boltzmann's constant.

\subsubsection{Virial expansion for the depletion potential}

The density functional formalism has been successfuly applied to calculate the 
depletion potential in HS mixtures \cite{rotheuro, rothpre}. 
Due to the lack of density 
functionals of proven adequacy for soft--sphere systems \cite{schmidt_sof_nota},
we need to implement an alternative approach.
One possibility is a systematic approach that provides a perturbative expansion for 
the depletion potential 
in powers of the solvent density. In the following we discuss such an expansion,
which has been discussed before and applied to HS mixtures \cite{Evans,almarza}. 

Within the second--order virial approximation, the grand potential 
associated to the solvent particles,
interacting via a pair potential $u_{22}(r)$ and in
the external potential $u_{12}(r)$ corresponding to
a colloidal particle located at the origin, is
\begin{eqnarray}
\beta \Omega[\rho_2]&=&\int d{\bf r}\rho_2({\bf r})\left[\ln\rho_2({\bf r})-1\right]
+\frac{1}{2}\int d{\bf r}\int d{\bf r}'\rho_2({\bf r})\rho_2({\bf r}')
f_{22}({\bf r}-{\bf r}')\nonumber\\
&+&\beta\int d{\bf r}\left[u_{12}({\bf r})-\mu_2\right]\rho_2({\bf r}),
\label{uno}
\end{eqnarray}
where $\rho_2({\bf r})$ is the density distribution of the solvent particles.
$f_{ij}=1-\exp{(-\beta u_{ij})}$ is the Mayer function
associated with the interaction between particles of species $i$ and $j$
(note that here it is defined with a negative sign with respect to the
standard definition). Within the same approximation, 
the chemical potential $\mu_2$ corresponding to a reservoir of solvent
particles with constant density $\rho_2$ is given by
\begin{eqnarray}
\beta\mu_2=\ln \rho_2+\rho_2\int d{\bf r}'f_{22}({\bf r}-{\bf r}').
\label{dos}
\end{eqnarray}
Functional minimization of Eqn. (\ref{uno}) with respect to $\rho_2({\bf r})$
and use of Eqn. (\ref{dos}) gives the following integral equation for the 
equilibrium density profile $\rho_2({\bf r})$:
\begin{eqnarray}
\rho_2({\bf r})=\rho_2\exp\left[-\beta u_{12}({\bf r})-
\int d{\bf r}'\left[\rho_2({\bf r}')-\rho_2\right]
f_{22}({\bf r}-{\bf r}')\right].
\label{tresa}
\end{eqnarray}
We will obtain the solution to Eqn. (\ref{tresa}) perturbatively, i.e. 
assuming the validity of the density expansion 
\begin{eqnarray}
\rho_2({\bf r})=\rho_2\Psi_1({\bf r})+\rho_2^2
\Psi_2({\bf r})+\cdots.
\label{cuatro}
\end{eqnarray}
The functions $\Psi_i({\bf r})$ are found by
inserting Eqn. (\ref{cuatro}) into Eqn. (\ref{tresa});
at first and second order in $\rho_2$ we find
\begin{eqnarray}
\Psi_1({\bf r})&=&\exp\left[
-\beta u_{12}({\bf r})\right] \label{found1},\\
\Psi_2({\bf r})&=&\exp\left[-\beta u_{12}({\bf r})\right]
\int d{\bf r}'f_{12}({\bf r}')f_{22}({\bf r}-{\bf r}').
\label{found2}
\end{eqnarray}
Up to third order in density, the contribution to the virial expansion of the 
excess part of the free--energy functional of the binary mixture
containing terms linear in 
$\rho_1({\bf r})$ is 
\begin{eqnarray}
\beta {\cal F}^{(1)}_{\rm{ex}}[\rho_1,\rho_2]=
\int d{\bf r}\int d{\bf r}' \rho_1({\bf r})\rho_2({\bf r}')f_{12}
({\bf r}-{\bf r}')
\nonumber\\
+\frac{1}{2}\int d{\bf r}\int d{\bf r}'\int d{\bf r}''
\rho_1({\bf r})\rho_2({\bf r}')\rho_2({\bf r}'')
f_{12}({\bf r}-{\bf r}')f_{12}({\bf r}-{\bf r}'')f_{22}
({\bf r}'-{\bf r}'').
\label{cinco}
\end{eqnarray}
Using the definition of $c_1^{(1)}({\bf r};[\rho_1,\rho_2])$, the one-body direct correlation 
function for big spheres, i.e. $c_1^{(1)}({\bf r};[\rho_1,\rho_2])
=-\delta\beta 
{\cal F}_{\rm{ex}}[\rho_1,\rho_2]/\delta \rho_1({\bf r})$, we find 
the dilute limit $\rho_1\to 0$ from Eqn. (\ref{cinco}) as
\begin{eqnarray}
&&-c^{(1)}_{1}({\bf r};[\rho_1=0;\rho_2])=
\int d{\bf r}'\rho_2({\bf r}')f_{12}({\bf r}-
{\bf r}')\nonumber \\
&&+\frac{1}{2}\int d{\bf r}'\int d{\bf r}''\rho_2({\bf r}')\rho_2({\bf r}'')
f_{12}({\bf r}-{\bf r}')f_{12}({\bf r}-{\bf r}'')
f_{22}({\bf r}'-{\bf r}'').
\label{seis}
\end{eqnarray}
The limit $\rho_1=0$ represents the infinite dilution limit of the mixture 
with respect to the big spheres. 
As shown in Ref. \cite{rotheuro, rothpre}, the depletion potential between 
two big particles in the sea of solvent particles can be 
obtained from 
\begin{eqnarray}
\beta W({\bf r})=c^{(1)}_{1}\left(\infty;[\rho_1=0,\rho_2]\right)-
c^{(1)}_{1}({\bf r};[\rho_1=0,\rho_2]). 
\label{siete}
\end{eqnarray}
Note that the calculation of the pair depletion potential 
requires the evaluation of the functional 
$c^{(1)}_{1}({\bf r};[\rho_1=0,\rho_2])$ at the equilibrium density profile of small particles
in the presence of only one big sphere, $\rho_2({\bf r})$, 
whose effect is taken care of by
an external potential. This feature 
makes density functional theory to be a powerful tool to calculate
solvation forces in multicomponent mixtures \cite{rotheuro, rothpre}.

Substituting Eqn. (\ref{seis}) into Eqn. (\ref{siete}) and using
Eqn. (\ref{cuatro}), with the expressions found for $\Psi_i({\bf r})$, 
Eqns. (\ref{found1}) and (\ref{found2}), we arrive finally at
\begin{eqnarray}
W({\bf r})&=& W^{(\rm{AO})}({\bf r})+W^{(\rm{B3})}({\bf r}),\label{total}\\
\beta W^{(\rm{AO})}({\bf r})&=& 
-\rho_2\int d{\bf r}'f_{12}({\bf r}')f_{12}({\bf r}
-{\bf r}'),\label{AO}\\
\beta W^{(\rm{B3})}({\bf r})&=&\rho_2^2
\int d{\bf r}'\int d{\bf r}'' f_{12}({\bf r}-{\bf r}')
f_{22}({\bf r}'-{\bf r}''){\cal K}({\bf r},{\bf r}',{\bf r}'') \label{B3},
\end{eqnarray}
where the kernel 
\begin{eqnarray}
{\cal K}({\bf r},{\bf r}',{\bf r}'')=e^{-\beta u_{12}({\bf r}')}
f_{12}({\bf r}'')+ \frac{1}{2}
\left\{
e^{-\beta[u_{12}({\bf r}')+u_{12}({\bf r}'')]}-1
\right\}f_{12}({\bf r}-{\bf r}'')
\label{follon}
\end{eqnarray}
has  been defined.
The Asakura-Oosawa (AO) approximation given by Eqn. (\ref{AO}), and the 
more accurate approximation given by Eqns. (\ref{total})--(\ref{follon}), which 
also includes a contribution from three-body interactions [see Eqn. (\ref{cinco})]
will be checked against computer simulations in Sec. \ref{Pres}.
All integrals 
involved in Eqns. (\ref{AO}) and (\ref{B3}) have been
evaluated numerically using Gaussian quadratures.
The above procedure can be systematically continued to obtain higher--order
contributions to the depletion potential. However, inclusion of the 
next term, proportional to $\left(\rho_2\right)^3$, requires 
the evaluation of multidimensional integrals with a numerical cost 
comparable to that required in MD simulations of the 
exact depletion potential. This is the reason why we stop the
density expansion at second order in the present work.

\begin{figure}
\includegraphics[scale=0.5]{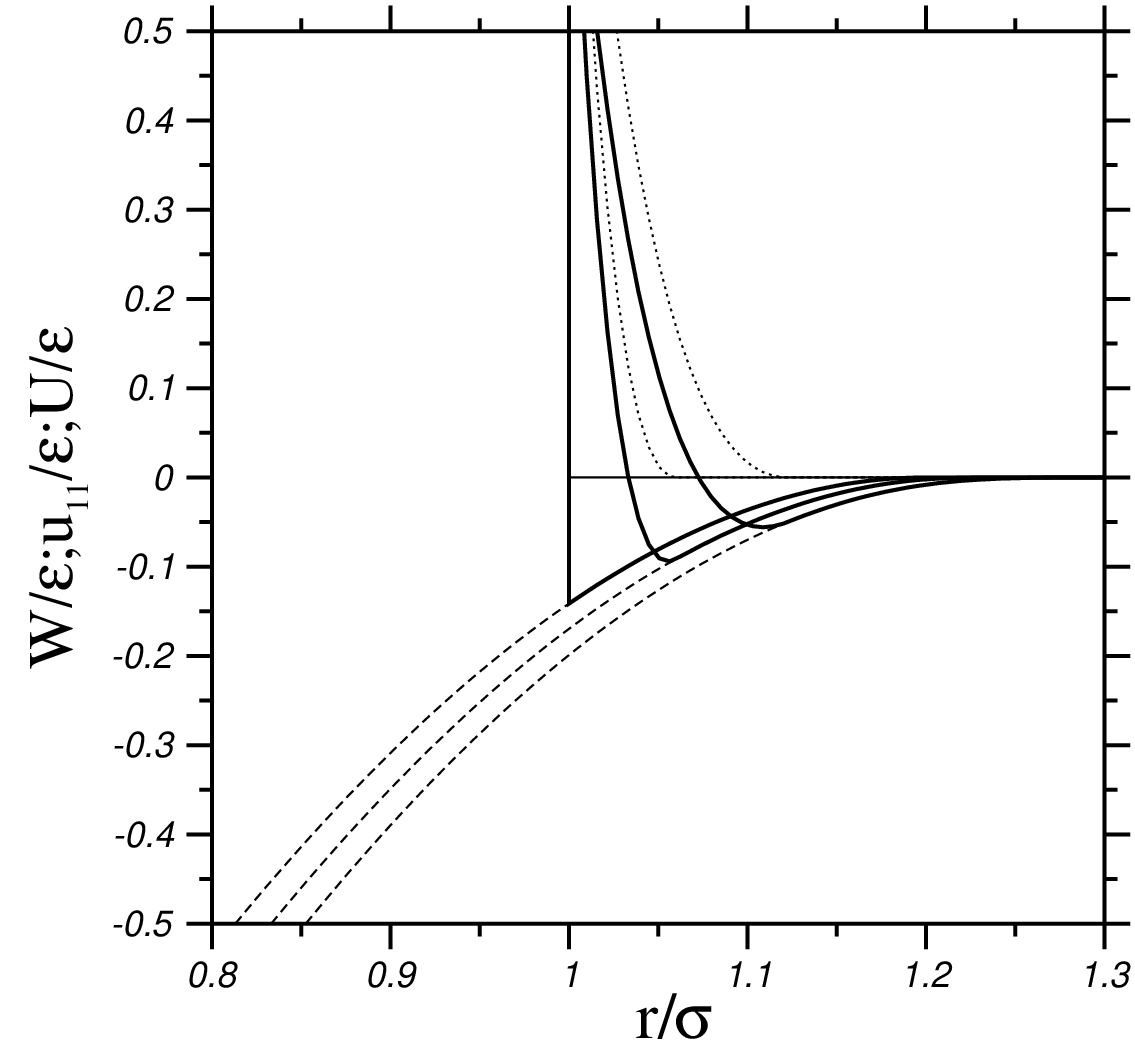}
\caption{}
\label{fig1}
\end{figure}

Let us now examine the predictions of the above two approximations with
respect to density $\rho_2^*$ and softness parameter $n$.
Fig. \ref{fig1} shows, for values of the softness parameter $n$=6, 12 and 
$\infty$, the depletion potential $W(r)$ evaluated with the AO approximation,
Eqn. (\ref{AO}), at a density $\rho_{2}^*=
0.0409$, sufficiently low that the use of this approximation could in principle
be justified, and at a temperature $T^*$=0.776.
The figure includes plots of the direct interaction $u_{11}(r)$
between two bigger particles, as well as the resulting
total interaction potential, $U(r)=u_{11}(r)+W(r)$,
obtained by the sum of direct and indirect contributions.
Significant differences can be seen among the attractive
depletion interactions as the softness parameter is changed: in particular,
the softer the pair interactions, the longer--ranged the attractive contribution.
As expected, the depletion potential in the case of the hard sphere
binary mixture vanishes at a distance $r^*\equiv r/\sigma=1.2$, while
$W(r)$ in the case of soft spheres does so at progressively 
larger distances. All depletion potentials follow the same trend, but
the softest direct potential, corresponding to the case $n=6$, exhibits a more
attractive indirect interaction. This effect is partly 
counterbalanced by the fact that in this case the direct interaction 
potential is softer, remaining positive at larger distances than in the other
cases (in fact, direct interactions vanish at a distance $r^*=2^{1/n}$, which 
increases as $n$ is reduced); the net effect is that the energy minimum
$U_{\rm min}$ of the total interaction potential deepens and shifts to shorter
distances as $n$ is increased. Now, a convenient measure of the well width 
is given by the quantity
$w_n$=$({{r}}^{(n)}_{2}-{{r}}^{(n)}_{1})/{{r}}^{(n)}_{1}$, with
$r^{(n)}_{i}$ being the distance between the origin and the $i$th root of 
$U(r)$. Using this criterion, it can be seen 
that the potential well becomes wider as the direct interactions 
operating in the mixture become softer. All these trends, which occur
at the level of the AO approximation, are monotonic
with the degree of softness.

\begin{figure}
\includegraphics[scale=0.5]{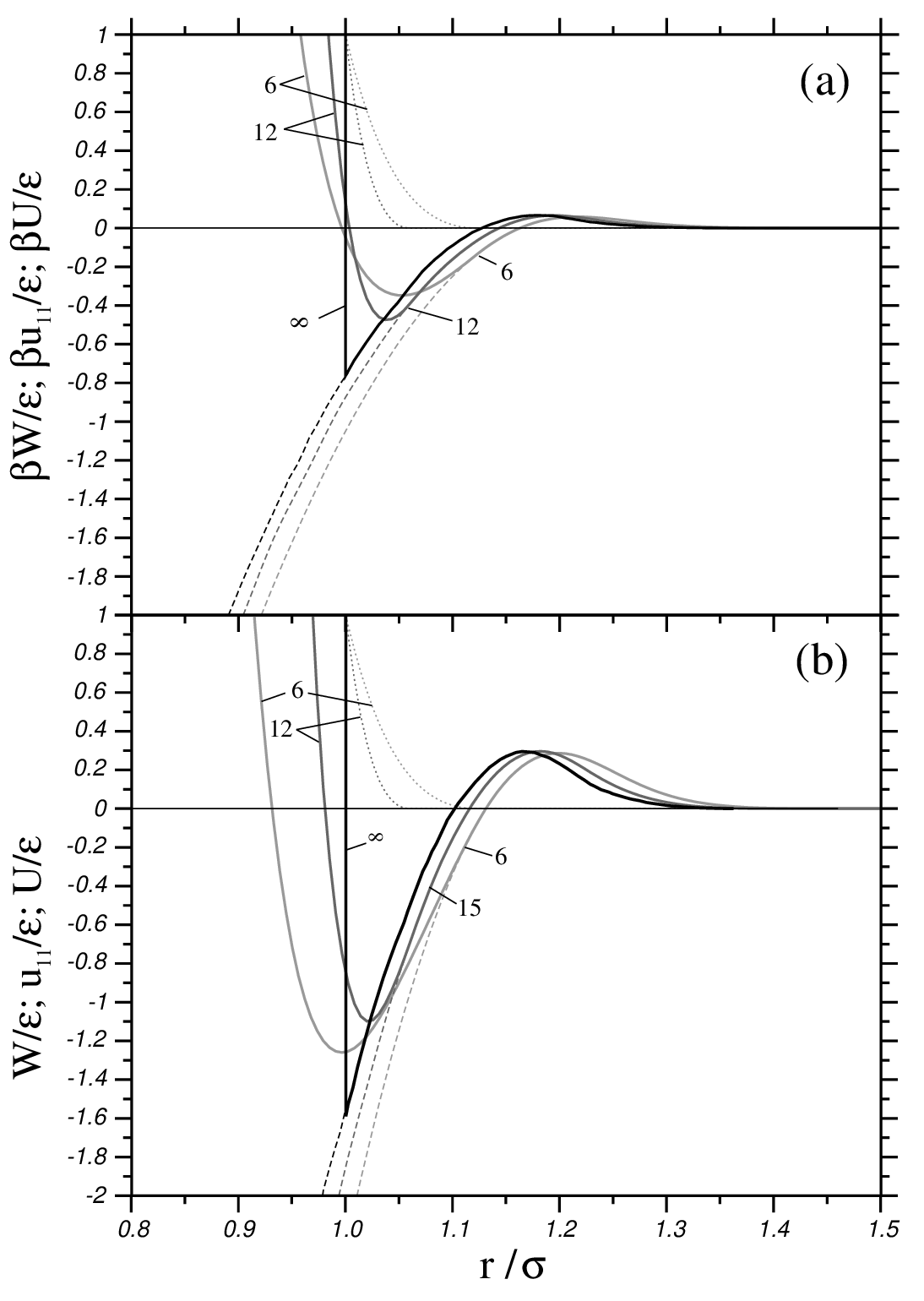}
\caption{}
\label{fig2}
\end{figure}

This situation changes to a certain extent when corrections from the second--order
density terms are incorporated, as illustrated in Fig. \ref{fig2}. Here depletion
potentials are calculated by employing Eqns. (\ref{total})--(\ref{follon})
at densities $\rho_{2}^{*}$= 0.2045 [Fig.\ref{fig2}(a)] and
$\rho_{2}^{*}$= 0.4090 [Fig. \ref{fig2}(b)], both at $T^*$=0.776.
As expected, the depletion potentials become more negative with density. 
In addition, they start to show maxima at distances $r^*\sim 1.15$--$1.20$; these
maxima become more pronounced as density is increased. The existence of 
an associated repulsive barrier in the depletion potentials is a 
well--established fact, first shown in Ref. \cite{walzsharma}.
The height of the maximum is rather insensitive to $n$,
while its location decreases very slightly with $n$.
However, by looking jointly at
Fig. 1 and 2, the combined action of $\rho_2^*$ and $n$ is noticeable.
For relatively low values of $\rho_2^*$, the smallest value of $U_{\rm min}$ 
is observed for $n\rightarrow \infty$, while $U_{\rm min}$ is a monotonically
decreasing function of $n$.
In the same density regime, the width of the potential energy well, $w_n$,
decreases steadily with $\rho_2^*$ at a fixed value of $n$; 
the same monotonically decreasing behavior of $w_n$ with $n$ occurs
even at higher densities, so that $w_6 > w_{12} > w_{\infty}$ always.
However, contrary to $w_{\infty}$, which persists in decreasing with density, 
$w_{12}$ at $\rho_{2}^{*}=0.4090$
is almost the same as that at $\rho_{2}^{*}=0.2045$,
whereas $w_6$ at the higher density is even larger than the value 
at $\rho_{2}^{*}=0.04090$. 
Finally, the value of $U_{min}$ for finite $n$ approaches that
for $n\rightarrow \infty$, still the most negative, but in a nonmonotonic
way, as the value of $U_{min}$ for $n=6$ is larger
than that for $n$=12.

Summing up to this point, we have shown that the density expansion of the
depletion potential up to second order indicates that the predictions of the
simple AO approximation, namely that the potential well associated to the 
total interaction shifts to longer distances and becomes less attractive 
with respect to the hard--sphere case as the direct interactions become 
softer, are different at higher fluid densities: the potential well 
and its depth behave nonmonotonically with the softness
parameter. A full calculation, valid to all orders in density, is needed
to elucidate whether these trends could give rise to enhanced attractive
interactions with respect to the hard--sphere mixture when the interactions
become sufficiently soft.

\subsubsection{Details on MD simulations}

The predictions presented in the previous section are based on a density expansion of $W(r)$.
Since it is not practical to improve our approximation beyond the second order
in density, one has to resort to computer simulation in order to check the validity of the results of the last section. We have carried out extensive MD
simulations on a system
of two big spheres immersed in a fluid of small spheres, calculating the depletion
potential directly; this technique allows us to obtain depletion potentials
which are, in principle, exact at all densities.

In the above theoretical derivation, it has been assumed from the beginning
that the two big spheres remained fixed at certain positions and that their
effect on the small spheres could be taken into account by means of an external 
potential. In our MD simulations, two methods have been used to freeze the 
dynamics of the big spheres.

In the first method, the distance, $r$, between the big spheres
has been maintained fixed in the course of the simulation run by using the 
method of constraints reviewed in Ref. \cite{cicco1}. Note that we assume
the two spheres to form a dimer, with translational and rotational 
degrees of freedom; it is only the dynamics associated to the
relative distance between the two big spheres which is frozen. 
This introduces an additional contribution to the free energy per particle with
respect to the case where the centers of mass of the two big spheres are fixed,
coming from the volume and solid angle explored by the dimer, but averaged 
quantities such as the potential of mean force are not affected. 
In the second method we fix the centers of mass of the big spheres,
which therefore act as an external potential on the small spheres.
In both methods, the force, $F(r)$, acting on these bigger particles
due to the smaller particles, has been calculated according to the formula
first used in Ref. \cite{cicco2} to evaluate the potential of mean force 
between two ions in a polar solvent, and already applied in the context of 
colloid physics in Ref. \cite{giapponesi}, that is:
\begin{equation}
F\left({r}\right)=\frac{1}{2}\left\langle \hat{\bm{r}}
\cdot \sum_{j=1}^{N_2}
\left(\bm{F}_{j}^{(a)}-\bm{F}_{j}^{(b)}\right) \right\rangle
\label{forza}
\end{equation}
In the above equation $\hat{\bm{r}}$ is the unit vector defining the
direction of the vector joining the two big spheres, $a$ and $b$, while
$\bm{F}_{j}^{(i)}$ is the force on the big sphere $i=a,b$ 
due to the small sphere 
$j$. The angular brackets indicate an average over the configurations 
of the small particles generated during the simulation run. This 
method cannot be applied in HS mixtures, which require evaluation of
the surface density distribution on the big spheres; possible inaccuracies 
introduced by extrapolation to contact of the density histogram
obtained from the simulation are not present here. Statistical
errors are of the same order, however, since in both, HS-- and soft--sphere
mixtures, the fluctuating quantity is averaged over $O(N_2)$ values.

Systems of two big particles, each of mass $m_1$, immersed in a sea of
$N_2$=2046 or 4000 small particles (depending on the method used to
calculate the depletion force), each one of mass $m_2=m_1$,
have been simulated with standard MD simulation techniques. Periodic
boundary conditions are applied on the rectangular simulation box.
The simulations have been carried out for the cases $n$=6 and 12, integrating
the equation of motion with the Verlet algorithm \cite{allentilde}, 
using a time step of 5$\times 10^{-4}$$\tau$,
$\tau$ being equal to $\sigma_{11}$$\left(m_{1}/\epsilon\right)^{1/2}$,
and maintaining the temperature
at the desired value via the Nos\'{e}--Hoover thermostat \cite{allentilde}. 
For each finite value of $n$ investigated, several distances have been 
considered, at which the depletion force is calculated during production
runs of $1$--$3\times 10^5$ time steps, preceded by $5\times 10^4$
time steps of equilibration. The depletion potential
is then obtained by numerically integrating this force. Before integration,
forces have been smoothed out by fitting to an appropriate function 
 or  by Akima spline approximation \cite{notasulfit}. The
integration constant has been set so that the depletion potential was chosen
to be zero at the largest radial distance considered.

The solvent density, $\rho_2^*$, which should correspond to the density of the
reservoir of solvent particles, is evaluated by computing the time--averaged
density profile at the boundaries, i.e. far from the big spheres, averaging 
over the boundaries to suppress fluctuations. This
density does not exactly coincide with that obtained by dividing
$N_2$ by the volume available to the small spheres, since the density
profile is highly structured in the region next to the big spheres.

Using the techniques described above, depletion interaction potentials, 
$W(r)$, and their corresponding total effective potentials, $U(r)$,
have been calculated for a few values of the density $\rho_2^*$, the
softness parameter $n$ and the temperature $T^*$. 

\section{Presentation and Discussion of the Results}
\label{Pres}

In this section we check the depletion potential resulting from the
density expansion proposed above, see Eqns. (\ref{total})--(\ref{follon}), against
our simulation results. The purpose of this exercise is to elucidate
whether a truncated density expansion is adequate to describe
depletion effects; in particular, we would like to estimate the maximum solvent density 
for which the truncated expansion is quantitatively valid. Then, the behavior of
the depletion potentials obtained from simulation with respect to density,
softness parameter and temperature will be assessed.

\subsection{Check of virial expansion against computer simulation}

First we compare the predictions obtained from the truncated virial
expansion for a fixed softness parameter, $n=6$, with the simulations
results, at fixed temperature $T^*$=1.000 and for several values of $\rho_2^*$.
Fig. \ref{depVSsimu} shows the results of this comparison. It appears that
the first--order expansion, corresponding to the AO theory, insufficiently
accounts for the simulation data even at the lowest density considered,
$\rho_2^*$ =0.190. The incorporation of the next term in the density expansion
improves significantly the performance of the theory with respect to simulation.
The agreement between the outcome of Eqn. (\ref{total}) and the simulation
data is very good for $\rho_2^*$ =0.190 at all distances. Increasing the density,
the agreement naturally deteriorates but is still quite fair at $\rho_2^*$ =0.381
for distances $0.80 < r^* < 1.15$. In contrast, the agreement between theory and 
simulation is poor at the very high density $\rho_2^*$ =0.740, which indicates
that higher--than--second order terms should be included in the density expansion.

Due to the ``compensatory'' effect of the addition of the direct interaction 
potential $u_{11}(r)$ at the shortest distances,
the agreement between the second--order virial expansion theory and simulation
is better if the total effective potentials are examined (Fig. \ref
{totVSsimu}).
For $\rho_2^*=0.190$, Eqn. (\ref{total}) predicts a potential $U(r)$ which is 
almost indistinguishable from that of simulation. At $\rho_2^* =0.381$,
the agreement between the two curves is still fairly good,
especially with regard to the location and width
of the attractive well, while the subsequent tiny maximum is 
displaced and its height slightly overestimated. It is only at larger values
of the density of the smaller spherical particles, such as $\rho_2^*=0.565$, that
the two potentials show noticeable discrepancies. For $\rho_2^* =0.740$,
the results disagree considerably even in the attractive region.

By confronting Eqns. (\ref{AO}) and (\ref{total}--\ref{follon}) together, and
by exploiting the results just presented, one can conclude
that Asakura--Oosawa theory can be used to construct
accurate total effective potentials up to density $\rho_2^* \simeq 0.05$;
the inclusion of the quadratic term is necessary
for densities in the range $0.05\alt \rho_2^* \alt 0.40$, while
for larger densities at least the cubic term is to be used.
Once we have assessed the density range where the virial expansion
is valid, we now turn to explore the dependence
of the depletion and total effective potentials obtained from
simulation with respect to density, the degree of softness and the
temperature.

\subsection{Dependence of the depletion potential on density, 
$n$ and temperature}

Fig. \ref{fig3} shows the values of the depletion force, calculated from MD simulation,
as a function of distance, for two values of the softness parameter,
$n=6$ and $n=12$, and for two high densities, $\rho_{2}^*$=0.680 and $\rho_{2}^*$=0.750; 
the temperature is $T^*=0.776$. Note that the values of density chosen 
are certainly below the freezing density of the WCA fluid, which accurate estimations
based on computer simulation locate at $\rho_2^*\simeq 0.91$ for $T^*=1.0$ \cite{Hess}.
In all cases the depletion force has been evaluated in the region of small
distances as well; note that $F(r)$ vanishes at $r=0$, i.e. when the two
big spheres are exactly superimposed, because of symmetry.
The computations of the depletion potential at such small 
distances could appear \emph{prima facie} useless,
as the direct interactions are so largely dominant in these conditions that
the value of $W(r)$ does not affect that of $U(r)$.
In particular, for hard--body systems, the evaluation of depletion forces
at distances shorter than the contact distance is clearly useless in the construction
of the effective pair potential. Indeed, previous simulation studies on
hard--sphere binary mixtures have only reported data at distances larger than the contact distance
\cite{bladon,dickmann}, and theoretical efforts have tried to reproduce these data
(\textit{e.g.} Ref. \cite{rotheuro, rothpre}). Nonetheless, indirect interactions are clearly
defined for all distances, regardless of the nature of the direct interactions, 
and their evaluation at very small distances should be of importance \textit{per se}
as they constitute a set of numerical data against which theories should be tested.

The shapes of the depletion forces shown in Fig. \ref{fig3} are clearly
consistent with two distinct regimes as the radial distance $r$ is changed: 
a shallow and deep minimum at scaled distances smaller than unity, and an oscillating part
at longer distances.
The value of the minimum is rather insensitive to the degree of softness but depends 
strongly on the value of density, while the location of the minimum is weakly $n$--dependent.
Changing these two parameters gives rise to rather different oscillatory structures, the
amplitude and damping of the latter being larger for increasing values of density and softness 
parameter.

As shown in Fig. 5, the depletion forces have been fitted, using empirical functions which
 reflect the existence of these two regimes \cite{notasulfit}.
The depletion potentials have been then obtained by integrating the corresponding
fitting function. The results are shown in Fig. \ref{fig4}, which contains
the direct interaction potentials together
with the resulting total effective pair potentials $U(r)$. Unfortunately,
simulation data on the analogous hard--sphere binary systems are available
only up to a density of $\rho_{2}^*=0.7$ \cite{biben,dickmann}. 
Thus, in order to make a comparison, the parametrized formula derived
in Ref.\cite{rothpre} for the depletion potential in hard--sphere mixtures has been
used at the highest density considered, $\rho_{2}^*=0.750$. 
%
%
From that figure, one can see that for the two densities considered, $\rho_{2}^*=0.680$ and
$0.750$, not only the well width but also the depth of the total
effective interactions between the big soft spheres are larger than in the hard--sphere case,
and in the case of the softest potential, $n=6$, the attractive interactions are
considerably enhanced. 

This result concurs with those of Ref. \cite{amokrane2} about the importance of taking into
account the direct interaction between the big particles. The total effective interactions
which govern the thermodynamical, structural and dynamical properties of these particle
systems, is in fact the sum of two contributions which may have, as in the present case, 
different signs but comparable absolute values in an interval of distances close to the 
range of the interaction between the big spheres. Thus, the final shape of the total effective pair 
potential is clearly very sensitive to the details of the direct and indirect contributions, 
both on the same footing. 

However, in contrast with the results in Ref. \cite{amokrane1}, where attractive
interactions were considered and depletion interactions of varying nature (attractive or
even repulsive) were obtained, in the case of repulsive direct interactions the depletion potentials
are always attractive and, at high fluid densities, greatly enhanced with respect
to those obtained for HS mixtures. In the light of these results, and considering that
in some cases HS mixtures exhibit a metastable fluid--fluid phase separation, 
one may wonder whether
enhanced attractive depletion forces may change the nature of demixing phenomena in
mixtures. This topic will be discussed in the next section. 

One last issue in the present subsection 
involves the dependence of the depletion potential on temperature, a
dependence that certainly exists as the model particles interact via smooth, as opposed
to hard, interactions. Therefore, the total interaction potential 
will in general be a parametric function not only of density, but also of temperature,
$U(r;\rho_2^*,T^*)$.
Fig. 7 shows the MD--derived depletion and total effective potentials 
obtained for
$n$=6 at densities $\rho_2^*=0.381$ and $0.740$, and temperatures
$T^*=0.776$ and $1.000$. As expected, density has a larger,
dominant effect. Differences between the two depletion potentials corresponding to 
the same density but different temperatures are discernible only at the shortest distances,
the two functions being essentially superimposed for $r^*>1$.
In the case $\rho_2^*=0.381$ the total interaction potentials are very close
even at short distances, where the two functions are positive.
When $\rho_2^*$=0.740 the difference is larger at these distances, but still small.
Irrespective of the value of density, the depletion potential becomes stronger as
temperature is increased, which is reflected in the total effective potential
having a deeper and wider attractive well; this can be understood in view
of the higher momentum interchanged between particles at higher temperatures,
which involves larger kinetic pressures (and hence larger depletion effects)
and more penetrable spheres.
In this respect, although density can be considered in the present context as the
more physically interesting variable, being responsible for the largest
variations in the effective potentials, the modest effect of temperature seen in 
Fig. \ref{effettoT} could have a large impact on the phase behavior and properties
of binary mixtures of the type investigated in this work.

\section{Calculation of the phase diagram for the repulsive soft--sphere mixture with $n$=6}

In the previous sections we have presented results for the depletion potentials of
mixtures of soft spheres at various densities of the small spheres. These potentials
have been computed, either using the virial expansion or by computer simulation,
under the assumption of infinite dilution of the big spheres.
Invoking the 
two--body approximation, we may assume that this is the potential felt by any
two pairs of big spheres when their density is arbitrarily large. Indeed, the effect 
of three-particle interactions on the depletion potential has been argued to
be overall very small \cite{bladon,pccp,gouldmelchpre,cinesi}.
In addition, we have made our own estimates of this 
effect by computing the force on a trimer of big soft spheres, in an equilateral triangular
configuration, at a distance between the spheres $r=\sigma_{11}+\sigma_{22}$,
which approximately
coincides with the maximum of the depletion potential, and then subtracting the
pair force associated to the three bonds; the remainder is the three-body 
contribution. Our calculation indicates that this contribution can be neglected altogether, 
confirming previous expectations \cite{bladon,pccp,gouldmelchpre,cinesi} in hard--sphere mixtures.

We now consider a collection of big spheres at density $\rho_1^*$ interacting 
via the depletion potential calculated at a density of small spheres $\rho_2^*$.
As mentioned in the introduction, a longstanding discussion was whether or not a similar 
mixture of hard spheres exhibits demixing, i.e. a first--order phase transition where two
phases with different concentration of big and small spheres coexist. The current
understanding is that hard--sphere mixtures do exhibit demixing,
but it involves a fluid phase and a crystalline phase, or two crystalline
phases \cite{dvre}. Since, as we have shown, soft repulsive spheres are influenced by a 
much stronger depletion force than in the case of hard spheres, it is pertinent
to ask the following questions: (i) Does a mixture of soft--repulsive spheres exhibit 
demixing? (ii) Is the tendency toward segregation stronger than in the hard--sphere 
case? (iii) Does demixing involve two fluid phases, rather than one or the two being 
crystalline? Since we only intend to explore the possible phase behaviors of the mixture,
we have applied a perturbation theory to calculate the
phase diagram; an alternative, in principle exact computer simulation approach
is for the moment ruled out because of its relatively high computational cost.
The theory is a density--functional theory based on the weighted--density approximation 
where a perturbation term is added, the so--called perturbative weighted density
approximation (PWDA). The perturbative term is evaluated using the compressibility
equation to fine--tune the effective radial distribution function in the crystal
phase. This procedure involves a mapping onto an effective HS fluid. 
Even though the theory cannot be trusted for liquid densities beyond the
crystallisation density for hard spheres, for densities less than that
the predictions of the theory should be qualitatively correct. 
More details on the theoretical procedure can be found in Ref. \cite{PWDA}. 

The resulting phase diagram, in the $\rho_1^*$--$\rho_2^*$ plane, is shown in Fig. 
\ref{PhaseDiagram}. The continuous lines correspond to the fluid--solid
binodal lines obtained from the
PWDA theory using depletion potentials calculated from the second--order
virial expansion, Eqn. (\ref{total}).
The dashed lines are the corresponding calculation of the spinodal lines
for fluid--fluid phase separation; note that the low-density spinodal
should be given qualitatively correctly by the PWDA theory (maybe not so
in the case of the high--density spinodal). It can be seen that fluid--fluid 
separation is always metastable
with respect to the fluid--solid transition. PWDA calculations based on depletion
potentials calculated from simulation are indicated by symbols. 
As expected, the two results are very similar for low values of $\rho_2^*$, where the 
truncated virial expansion
is supposed to be accurate. At higher densities $\rho_2^*$, the two sets of depletion
potentials give different but compatible results (for both binodal and
spinodal lines). 

For the sake of illustration we
include in Fig. \ref{PhaseDiagram}, as dotted lines, the results for the
fluid--solid binodal line from a
slightly different version of perturbation theory \cite{PRE}
as applied to the corresponding HS mixture using the depletion potential 
of Roth et al. \cite{rothpre}. In fact, the HS fluid--fluid spinodal line
predicted by this theory occurs at very high solvent densities, too high
to appear in the range of Fig. \ref{PhaseDiagram}; given that
the effective depletion potential is not valid in this regime \cite{rothpre}, we 
conclude that, for the present value of size ratio, fluid--fluid phase separation 
in effective HS mixtures is, if anything, highly metastable with respect to
direct fluid--solid phase equilibrium. In contrast, in the case of soft
interactions, fluid--fluid phase separation is close to being stable
with respect to fluid--solid phase separation. 
The predictions of perturbation theory as to the large differences 
between HS and soft--sphere mixtures is very significant
and illustrate the effect on the 
phase behavior of the significantly more attractive effective interactions 
operating in soft-sphere mixtures than in the corresponding HS mixtures. 

One can speculate
on possible variations of the soft repulsive bare potentials (in the direction of
making them even softer, for instance) that could promote the stability of the 
fluid phase, and its associated critical point, with respect to the solid phase.
Actually, recent work \cite{tejero,hanegawa} supports the possibility that an increasingly
deeper and, especially, wider attractive well could lead to the observation of
a stable fluid--fluid phase separation in these binary systems. In any case, use
of an improved theory and more extensive simulations on fluids with other types of
interactions and at different temperatures are needed, since the final outcome appears 
to be the result of a delicate free-energy balance between the fluid and the solid. 
Further exploration of these topics are left for future work.

\section{Concluding remarks}
\label{III}

The present results are considered pertinent
to real experimental binary colloidal dispersions. It is known that many of them
exhibit a fluid--fluid demixing transition and a phase behavior which is temperature
dependent (for a discussion of the
effect of temperature on the properties of
sterically stabilized colloidal systems see Ref. \cite{moises}),
two facts that are not captured by assimilating the
colloidal particles to hard bodies but that could be by modeling them
as soft--core objects.

Indeed, the pair--potential models investigated should be relevant to sterically
stabilized colloids. In these systems, the degree of softness of the direct
interactions can be controlled by varying the chemical composition and/or,
especially, the thickness of the coating polymer layer:
we could in principle expect that an increase in thickness 
 produces softer
interactions. Recently, an experimental study has been carried out on binary
dispersions formed by colloids sterically stabilized with a coating
polymer layer of varying chemical composition and thickness \cite{duijneveldt}.
Although no stable fluid--fluid separation was observed, evidence for weakly
metastable fluid--fluid demixing has been suggested for the mixture
with the thickest stabilizing layer. These
results have been interpreted in terms of a non--additive hard--sphere model.
However, another, perhaps physically more sound, explanation of these
experimental facts can be given in terms of the degree of interaction softness.
An interesting sequel of the study of Hennequin et al. \cite{duijneveldt}
could be an investigation of additional binary suspensions of sterically stabilized
colloids with varying thicknesses of the coating polymer layer. Further
theoretical work along the lines presented in the previous section could also
be valuable and guide possible experimental research avenues.

We hope that the present results will stimulate the search for real colloidal 
binary mixtures where interparticle interactions can be tuned from hard-- to 
soft--sphere--like, as well as an investigation of how the phase behavior and the 
properties of such mixtures evolve upon changing such characteristics.
Indeed, charge-- and sterically stabilized monodisperse colloidal suspensions
have recently been reported where the softness of the interactions is controlled
by the concentration of the added salt \cite{yethiraj}. Although the pair potential models
investigated in this work are not appropriate for charge--stabilized colloids,
as the presence of charges requires a different functional form for the model pair
interactions, the observed evolution of the depletion and
total effective interactions with density and degree of softness of
the direct interactions should be quite general. Thus, we believe it would be 
interesting to extend the study of Ref. \cite{yethiraj} to bidisperse suspensions.

\acknowledgments
MIUR (Italy) and Ministerio de Educaci\'{o}n y Ciencia (Spain) are thanked 
for financial support under the 2005 binational integrated program.
Y.M.-R. was supported by a Ram\'on y Cajal research contract.
This work is part of the research
projects PRIN "Energy and Charge Transfers
at molecular level"  of MIUR (Italy), MOSAICO, FIS2005-05243-C02-01 and FIS2004-05035-C03-02
of the Ministerio de Educaci\'on y Ciencia (Spain), and S-0505/ESP-0299 of
Comunidad Aut\'onoma de Madrid (Spain).

\clearpage

\textbf{FIGURE CAPTIONS}
\vspace{1cm}

Figure 1: Dashed lines: depletion potentials calculated with Eqn. (\ref{AO}) for $n=6$,
$n=12$ and $n\rightarrow \infty$ (from bottom to top).
Dotted lines: direct potential for $n=6$ (upper curve) and $n=12$ (lower curve);
the case $n\rightarrow \infty$ is a vertical line at $r=\sigma$.
Continuous thick lines: from right to left, total effective potentials
for $n=6$, $n=12$ and $n\rightarrow \infty$. All calculations
for a density $\rho_{2}^*= 0.0409$ and temperature $T^*$=0.776.
\vspace{1cm}

Figure 2: Dashed lines: depletion potentials calculated with Eqns. (\ref{total})--(\ref{follon})
for $n=6$, $n=12$ and $n\rightarrow \infty$.
Dotted lines: direct potential for $n=6$ (light gray) and $n=12$ (gray);
the case $n\rightarrow \infty$ is a vertical line at $r=\sigma$.
Continuous thick lines: total effective potentials, corresponding to softness
parameters $n=6, 12$ and $\infty$ indicated by labels and different gray
intensity. (a) $\rho_{2}^*=0.2045$ and (b) $\rho_{2}^*=0.4090$. In both cases $T^*$=0.776.
\vspace{1cm}

Figure 3: Depletion potential for $n$=6 calculated using Eqn. (\ref{AO})
(dotted lines), Eqns. (\ref{total})--(\ref{follon}) (dashed lines), and MD simulations
(solid lines) at densities $\rho_2^*$ =0.190 (a), 0.381 (b) and 0.740 (c). In all
cases $T^*=1.000$.
\vspace{1cm}

Figure 4: Total effective potentials for $n$=6 calculated using
Eqns. (\ref{total})--(\ref{follon}) (dashed lines) and MD simulations (solid lines)
at densities $\rho_2^*$ =0.190 (a), 0.381 (b), 0.565 (c) and 0.740 (d). In all
cases $T^*=1.000$.
\vspace{1cm}

Figure 5: The depletion force as a function of distance for:
$n$=6, $\rho_2^*$=0.680 (a);
$n$=6, $\rho_2^*$=0.750 (b);
$n$=12, $\rho_2^*$=0.680 (c);
$n$=12, $\rho_2^*$=0.750 (d).
Dots are the simulation results while lines are fitting curves
\cite{notasulfit}. In all cases $T^*=0.776$. Insets are zooms of the regions
about the maxima.
\vspace{1cm}

Figure 6: Dashed lines: depletion potentials obtained by integrating the
depletion force calculated in the simulations for $n=6$ and $12$
(value of $n$ indicated as a label).
Dotted lines: direct potentials for $n=6$ and $12$ (the case $n=6$
corresponds to the softer potential).
Solid lines: total effective potentials for $n=6$, $12$ and $\infty$
(HS), indicated by labels and different gray intensities (light, medium and
dark, respectively).
 Results in
(a) are for $\rho_2^*=0.680$, while those in (b) are for
$\rho_2^*=0.750$.
Data for HS mixtures ($n=\infty$)
from simulations of Ref. \cite{dickmann} (a), and
from density functional theory of Ref. \cite{rothpre} (b). Insets are enlargements of the regions
about the maxima.
\vspace{1cm}

Figure 7: Effect of temperature on depletion (squares)
and total effective (circles) interactions between two big spherical particles: (a)
$\rho_{2}^{*}$=0.381; (b) $\rho_{2}^{*}$=0.740. Open symbols correspond to
$T^*=0.776$, while filled circles were calculated with $T^*=1.000$. All results are for $n$=6.
\vspace{1cm}

Figure 8: Phase diagram for the effective one--component system in the plane
$\rho_1^*$--$\rho_2^*$. Solid lines: fluid--solid
binodal lines, as predicted by PWDA theory using virial approximation for
depletion potential, Eqns. (\ref{total})--(\ref{follon});
dashed lines: fluid--fluid spinodal lines as obtained from the same
theory. Dotted lines: fluid--solid binodals for a hard--sphere mixture
of the same size ratio,
as obtained from perturbation theory (see Ref. \cite{PRE}).
Filled circles: fluid--solid phase boundaries as obtained from PWDA theory
using depletion potential calculated from simulation. Open circles:
fluid--fluid spinodal from the same theory. Filled squares:
fluid--solid coexistence densities for the one--component WCA system
from computer simulation (see Ref. \cite{Hess}).
The value of the reduced temperature is $T^*=1.000$.
\vspace{1cm}

\clearpage

\clearpage

\clearpage
\begin{figure}
\includegraphics[scale=0.7]{fig3.eps}
\caption{}
\label{depVSsimu}
\end{figure}

\clearpage
\begin{figure}
\includegraphics[scale=0.6]{fig4.eps}
\caption{}
\label{totVSsimu}
\end{figure}

\clearpage
\begin{figure}
\includegraphics[scale=1.0]{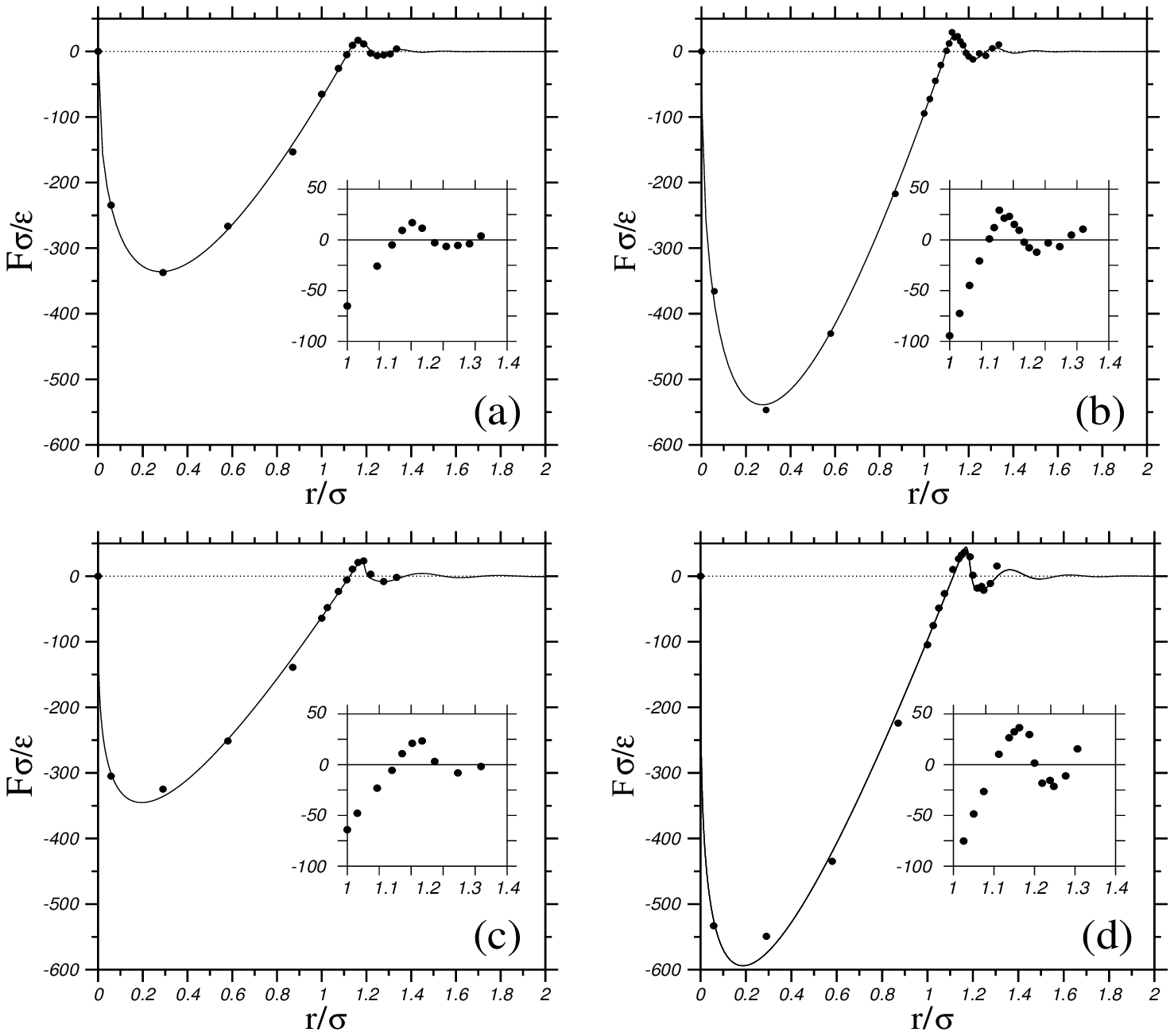}
\caption{}
\label{fig3}
\end{figure}

\clearpage
\begin{figure}
\includegraphics[scale=0.5]{fig6.eps}
\caption{}
\label{fig4}
\end{figure}

\clearpage
\begin{figure}
\includegraphics[scale=0.5]{fig7.eps}
\caption{}
\label{effettoT}
\end{figure}

\clearpage
\begin{figure}
\includegraphics[scale=0.38]{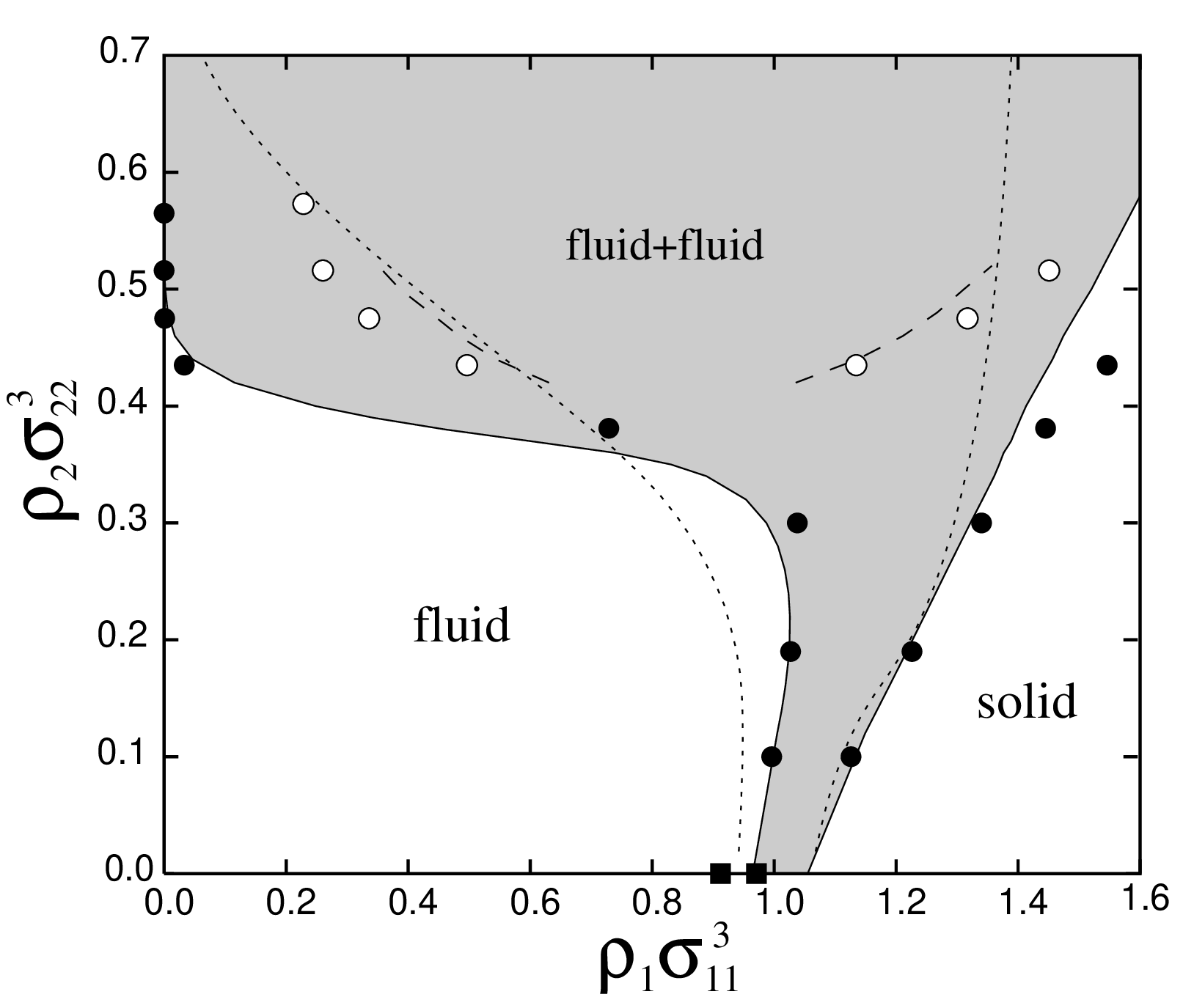}
\caption{}
\label{PhaseDiagram}
\end{figure}

\begin{thebibliography}{13}
\bibitem{schmidt_sof_nota} One  interesting, but in the present context
yet untested, possibility could be the density functional theory
developed in Ref. \cite{schmidt_sof}.
\bibitem{notasulfit}
The shape of the depletion force suggests to fit
the numerical data displayed in Fig. \ref{fig3} with the following empirical
function:
\begin{equation}
F^{\rm{fit}}\left(r\right)=
A\left[1-f(\alpha r^{\nu})\right]
\left(r^{\beta}+a r\right) 
- b f(\alpha r^{\nu})e^{-\kappa r}
\cos{\left[2\pi \delta \left(r-r_0\right)\right]}\nonumber
\end{equation}
with $\{A,\alpha,\nu,\beta,a,b,\kappa,\delta,r_0\}$ the set of fitting parameters.
Clearly this equation does not have any ambition to be a proper
description of the functional dependence of depletion forces upon distance; it
has just been used to smooth the numerical data.
In the equation one can recognize two regimes:
the attractive well described by the term $r^{\beta}+a r$, and the
subsequent oscillatory damped part described by the exponential--sinusoidal term.
The factor $f(\alpha r^{\nu})$ and its complement to unity
have the role of switching continuously from one regime to the other. 
The above functional form has proven to be quite effective
for $n$=6, and not so much for $n$=12 (see Fig. \ref{forza}); actually, for 
the computation of the potential $W(r)$ by integrating the force, the Akima spline 
method was used instead.
\end{thebibliography}
\end{document}